\documentclass{article}

\usepackage{graphicx} 
\usepackage{times}

\usepackage[final]{EMNLP2023}

\usepackage{comment}
\usepackage[utf8]{inputenc} 
\usepackage[T1]{fontenc}    
\usepackage{hyperref}       
\usepackage{url}            
\usepackage{booktabs}       
\usepackage{amsfonts}       
\usepackage{nicefrac}       
\usepackage{microtype}      
\usepackage{xcolor}         
\usepackage{multirow} 
\usepackage{graphicx}
\usepackage{booktabs}
\usepackage{tabularx}

\newcommand{\modify}[1]{{#1}}

\definecolor{mblue}{RGB}{0, 61, 124}
\definecolor{myellow}{RGB}{239, 124, 0}
\definecolor{mnavy}{RGB}{0,0,128}
\definecolor{minc}{RGB}{0,128,0}
\definecolor{mdec}{RGB}{255,0,0}
\definecolor{mhold}{RGB}{128,128,128}
\newcommand{\stitle}[1]{\vspace{1mm} \noindent {\bf #1}}

\newcommand{\equal}{\textsuperscript{*}Equal Contribution, Ordered Alphabetically}
\newcommand{\ms}[2]{{#1\tiny{$\pm$#2}}}

\title{A Reflective LLM-based Agent to Guide Zero-shot Cryptocurrency Trading}

\author{Yuan Li\thanks{\equal}, Bingqiao Luo\footnotemark[1], Qian Wang\footnotemark[1], Nuo Chen, Xu Liu, Bingsheng He \\
National University of Singapore\\
\texttt{li.yuan@u.nus.edu}, \texttt{luo.bingqiao@u.nus.edu}, \texttt{qiansoc@nus.edu.sg} \\ 
\texttt{nuochen@comp.nus.edu.sg},
\texttt{liuxu@comp.nus.edu.sg} \\
\texttt{hebs@comp.nus.edu.sg}\\
}

\begin{document}

\maketitle

\begin{abstract}
The utilization of Large Language Models (LLMs) in financial trading has primarily been concentrated within the stock market, aiding in economic and financial decisions. Yet, the unique opportunities presented by the cryptocurrency market, noted for its on-chain data's transparency and the critical influence of off-chain signals like news, remain largely untapped by LLMs. This work aims to bridge the gap by developing an LLM-based trading agent, CryptoTrade, which uniquely combines the analysis of on-chain and off-chain data. This approach leverages the transparency and immutability of on-chain data, as well as the timeliness and influence of off-chain signals, providing a comprehensive overview of the cryptocurrency market.  CryptoTrade incorporates a reflective mechanism specifically engineered to refine its daily trading decisions by analyzing the outcomes of prior trading decisions. This research makes two significant contributions. Firstly, it broadens the applicability of LLMs to the domain of cryptocurrency trading. Secondly, it establishes a benchmark for cryptocurrency trading strategies. Through extensive experiments, CryptoTrade has demonstrated superior performance in maximizing returns compared to traditional trading strategies and time-series baselines across various cryptocurrencies and market conditions. Our code and data are available at \url{https://anonymous.4open.science/r/CryptoTrade-Public-92FC/}.
\end{abstract}

\section{Introduction}
Large Language Models (LLMs) have revolutionized financial decision-making and stock market prediction by excelling in tasks such as sentiment analysis \cite{liang2022holistic} and explanation generation \cite{pu2023summarization}. Specialized models like FinGPT and BloombergGPT \cite{liu2023fingpt, wu2023bloomberggpt} demonstrate this capability. Recent research highlights their ability to interpret financial time-series and enhance cross-sequence reasoning \cite{wei2022chain, yu2023temporal, zhang2023multimodal, zhao2023survey, yang2024harnessing}. Furthermore, the development of LLM-based trading agents like Sociodojo \cite{cheng2024sociodojo} underscores the potential for innovating investment strategies.

\begin{figure*}[ht]
\centering 
\includegraphics[width=1\linewidth]{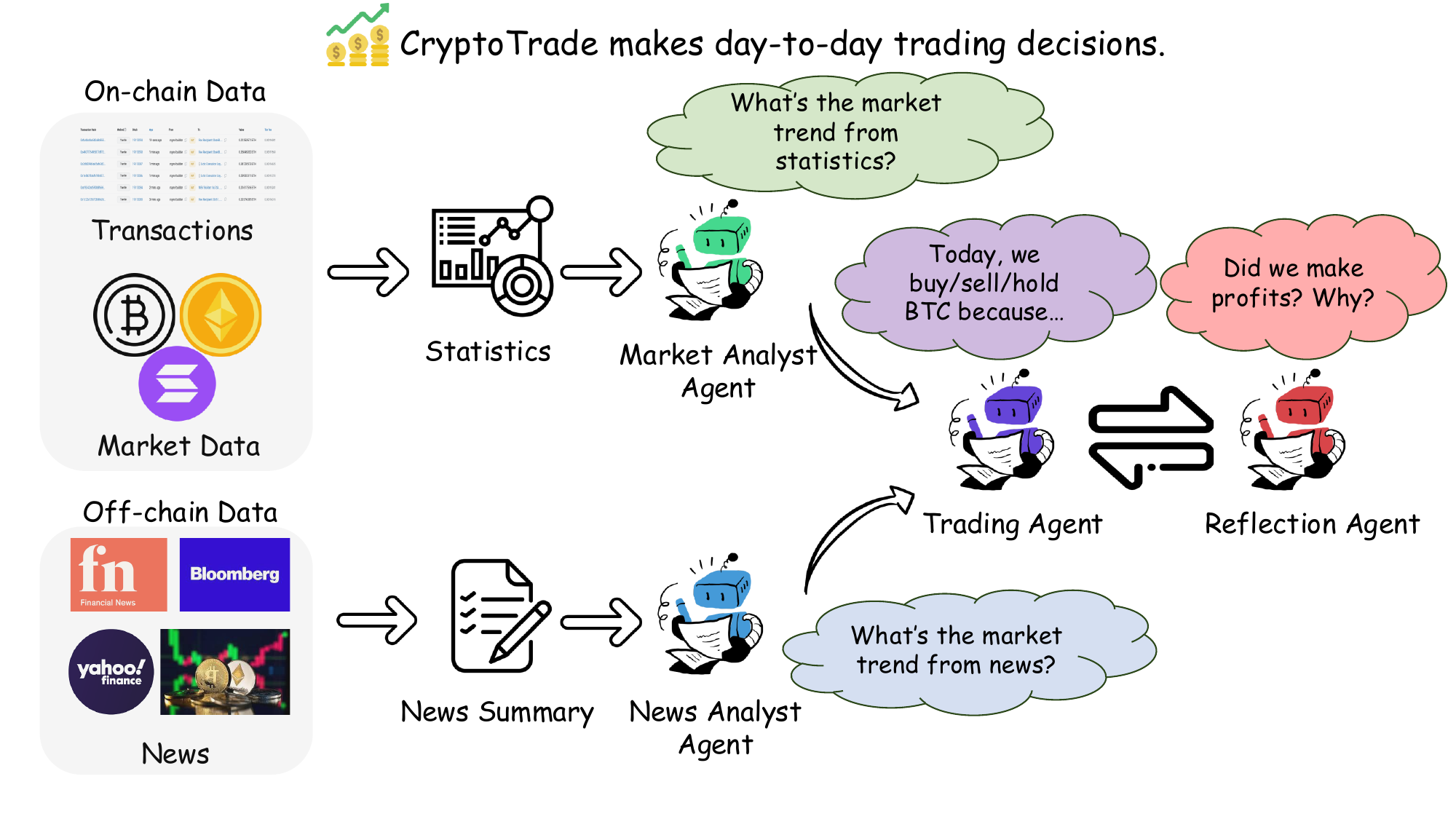} 
\caption{CryptoTrade Framework. Our framework begins with the collection of various types of data, including on-chain transactions, market data, and off-chain data from multiple financial news sources. We extract on-chain statistics while summarizing off-chain news to provide comprehensive inputs for our agents' analysis. We then deploy several LLM-based agents to make day-to-day trading decisions, utilizing a reflective mechanism to maximize total returns over different time periods.} 
\label{fig:overview} 
\end{figure*}

However, the application of LLMs in the cryptocurrency market remains underexplored, yet this field holds great potential for future development for three main reasons. First, the cryptocurrency market is characterized by high market value, volatility, and uncertainty, which challenge traditional trading signals \cite{drozdz2023mature, wei2023cryptocurrency}. Second, LLMs have demonstrated their ability to understand and analyze financial markets by leveraging large volumes of multi-modal data, such as news and price information \cite{wu2023bloomberggpt}. Third, the cryptocurrency market includes open-sourced on-chain data, such as gas prices and total transaction values, providing insights beyond just price movements \cite{feichtinger2023hidden, ren2023interoperability}. To bridge this gap, we introduce CryptoTrade. By integrating on-chain data, including market data and transaction records, with off-chain information like financial news, CryptoTrade leverages both dimensions to execute daily trading strategies, taking full advantage of the transparency of on-chain data and the immediacy of off-chain information. We detail the structure of CryptoTrade in \autoref{fig:overview}.


CryptoTrade consists of a three-part framework. Initially, we collect data where on-chain details such as transactions and broader market data are aggregated alongside off-chain data from established financial news outlets like Bloomberg and Yahoo Finance. Following collection, we conduct statistical analyses to derive indicators such as moving averages and utilizing text processing for news summarization using  GPT-3.5-turbo\footnote{\url{https://platform.openai.com/docs/models/gpt-3-5-turbo}}. Finally, we enhance day-to-day decision-making with specialized analytical agents: market analyst agent evaluates market trends, news analyst agent interprets recent news impacts, and trading agent deliberates on investment actions. Concurrently, reflection agent reviews past performance, allowing CryptoTrade to refine its strategies to maximize returns. 

Then, we conduct comprehensive experiments with CryptoTrade using GPT-4\footnote{\url{https://platform.openai.com/docs/models/gpt-4}}, and GPT-4o\footnote{\url{https://platform.openai.com/docs/models/gpt-4o}}, evaluating its proficiency in making daily trading decisions for Bitcoin (BTC), Ethereum (ETH), and Solana (SOL). These three cryptocurrencies were selected for their prominence and market values of \$134.14, \$45.59, and \$7.61 billion, respectively, as of June 2nd, 2024\footnote{\url{https://coinmarketcap.com/currencies/}}. CryptoTrade significantly outperforms time-series baselines such as Informer \cite{zhou2021informer} and PatchTST \cite{nie2022time}, and achieves comparable performance to trading signals like Moving Average Convergence Divergence (MACD) \cite{gencay1996non} in both return and sharpe ratio across bull, sideways, and bear market conditions. Notably, CryptoTrade operates in a zero-shot manner without fine-tuning based on validation sets, highlighting its potential for future applications. For instance, during the ETH bullish test period, the Buy and Hold strategy secured a 22.59\% return, while CryptoTrade exceeded this by a remarkable 3\%.

To summarize, we make the following three contributions:
\begin{itemize}
    \item We introduce CryptoTrade, an innovative trading agent in the cryptocurrency domain, driven by LLMs. CryptoTrade is designed to generate optimized trading decisions specifically for the cryptocurrency market, setting a new benchmark in this field.
    
    \item We develop a comprehensive framework for cryptocurrency trading agents that encompasses the collection of both on-chain and off-chain data, along with the integration of a self-reflective component to enhance decision-making processes. This approach aggregates diverse information sources and establishes a new standard for data-driven trading strategies within the cryptocurrency domain.
    
    \item Through rigorous experiments, we present empirical evidence showcasing the efficacy of CryptoTrade compared to other baselines. CryptoTrade advances the frontier of cryptocurrency trading technologies and offers valuable insights for financial decision-making.
    
\end{itemize}

\section{CryptoTrade Framework}

This section details the components employed to develop the CryptoTrade agent, including data collection, market dynamics analysis, and agents development. \autoref{fig:overview} shows an overview of CryptoTrade.

\subsection{Data Collection}
The foundation of our methodology relies on a comprehensive collection of data from on-chain and off-chain sources, which is essential for making informed trading decisions in the cryptocurrency market. The data license is detailed in \autoref{license}. The data ethics are explained in \autoref{collection.ethics}. The data collection strategy is illustrated in \autoref{fig:overview}(a) and further detailed below:

\begin{itemize}
\item \textbf{On-chain Data:} We leverage historical data from CoinMarketCap\footnote{\url{https://coinmarketcap.com}}, which provides daily insights into prices, trading volumes, and market capitalization of various cryptocurrencies: BTC, ETH, SOL. This dataset forms the backbone of our market trend analysis, enabling us to decipher long-term trends and identify cycles in cryptocurrency valuations and investor behavior.

Additionally, we incorporate detailed transaction statistics from on-chain activities. All blockchain transactions are transparent, traceable, and publicly accessible, achieved through securely linked blocks using cryptographic techniques \cite{narayanan2016bitcoin}. As numerous prominent blockchain explorers provide tools for easy access to blockchain transaction data, we retrieve on-chain transaction data from the Dune Database\footnote{\url{https://dune.com/home}}, a crypto analytics platform, and construct comprehensive statistics related to these transactions to include information on market dynamics. This includes comprehensive metrics such as daily number of transactions, number of active wallet, total value transferred, average gas price, and total gas consumed. These features are crucial for understanding the operational aspects of blockchains, such as network congestion times and cost efficiency, which directly impact trading strategies. Our daily collection of these metrics facilitates a nuanced analysis of market dynamics and liquidity, allowing for real-time adjustments to our trading algorithms based on current market conditions.

\item \textbf{Off-chain Data:} We employ the Gnews API\footnote{\url{https://pypi.org/project/gnews/}} to systematically gather news articles related to each cryptocurrency. This tool enables us to access a wide array of sources through Google News, providing a comprehensive daily snapshot of market sentiment. Moreover, we particularly focus on filtering news from reputable financial and cryptocurrency-specific outlets such as Bloomberg, Yahoo Finance, and crypto.news\footnote{\url{https://crypto.news/}} to ensure the reliability and relevance of the information. The integration of analysis from these articles allows us to capture the market's sentiment and response to developments, which is often a precursor to significant market movements.
\end{itemize}

By merging both on-chain data and off-chain news insights, our methodology offers a holistic view of the cryptocurrency market. This integration not only enhances our analytical capabilities but also significantly improves the precision of our trading decisions.

\subsection{Market and News Analyst Agents}
Upon collecting extensive on-chain and off-chain data, we analyze it through two key components of our CryptoTrade agent: (1) market analyst agent, (2) news analyst. By leveraging the capabilities of GPT-3.5-turbo, these analysts provide deep insights into the crypto market, enabling informed and strategic trading decisions.

\textbf{Market Analyst Agent.} The market analyst agent plays a crucial role in deciphering market dynamics through the statistical analysis of key trading signals from on-chain data, such as MA \cite{gencay1996non}, MACD \cite{wang2018predicting}, and Bollinger Bands \cite{day2023profitability}. Details of these trading signals are provided in \autoref{baselines}. Armed with this information, the market analyst agent compiles reports on the market's direction and momentum. An example is shown in \autoref{fig:market}.

\textbf{News Analyst Agent.} The news analyst agent is tasked with extracting and analyzing critical information from the latest news to assess the potential market impact of off-chain social hype. By sourcing news summaries from various trusted sources, the news analyst agent pinpoints relevant recent events and assesses the significance and implications of key topics, thus adding an extra dimension of insight. An example is provided in \autoref{fig:news}.

\subsection{Trading Agent}
Each day, the trading agent offers an investment suggestion based on reports from the market and news analyst agents. After analyzing the reports, the trading agent provides a concise rationale for its decisions. It also recommends allocating a certain portion of remaining cash to purchase cryptocurrency (with a range from $($\(0\) to \(1]\)), selling a certain portion of owned cryptocurrency (with a range from \([-1\) to \(0)\)), or holding (neither buying nor selling). When a trading decision is made, a transaction fee is charged in proportion to the traded value.
\autoref{fig:investment} illustrates an example of our trading agent's operations.

\subsection{Reflection Agent} 
The reflection agent reviews the trading agent's recent activities to enhance future strategies. By analyzing the previous week's prompts, decisions, and returns, the reflection agent identifies the most impactful information and the reasons behind its significance, providing feedback to the trading agent for future decisions. Consequently, CryptoTrade learns to focus on the most influential information for upcoming decisions. An example is illustrated in \autoref{fig:reflection}.

\section{Experiments}

In this section, we detail the experiments designed to evaluate the efficacy of our proprietary CryptoTrade agent in comparison to established baseline strategies within the trading domain. 

\subsection{Experimental Setup}
\textbf{Experiment Environments. }We conduct all experiments using PyTorch on an NVIDIA GeForce RTX 3090 GPU. More details are in \autoref{env}.


\begin{table*}[h!]
    \centering
\caption{Dataset splits. Prices are in US dollars. In each split, the transaction days include the start date and exclude the end date. We evaluate the total profit on the end date.}
\begin{tabular}{clccrrr}
\hline
Type                 & Split           & Start      & End        & Open     & Close    & Trend    \\ \hline
\multirow{4}{*}{BTC} & Validation      & 2023-01-19 & 2023-03-13 & 20977.48 & 20628.03 & \textcolor{mhold}{-1.67\%}  \\
                     & Test   Bearish  & 2023-04-12 & 2023-06-16 & 30462.48 & 25575.28 & \textcolor{mdec}{-15.61\%} \\
                     & Test   Sideways & 2023-06-17 & 2023-08-25 & 26328.68 & 26163.68 & \textcolor{mhold}{-0.83\%}   \\
                     & Test   Bullish  & 2023-10-01 & 2023-12-01 & 26967.40 & 37718.01 & \textcolor{minc}{39.66\%}  \\ \hline
\multirow{4}{*}{ETH} & Validation      & 2023-01-13 & 2023-03-12 & 1417.13  & 1429.60  & \textcolor{mhold}{0.88\%}   \\
                     & Test   Bearish  & 2023-04-12 & 2023-06-16 & 1892.94  & 1664.98  & \textcolor{mdec}{-12.24\%} \\
                     & Test   Sideways & 2023-06-20 & 2023-08-31 & 1734.79  & 1705.11  & \textcolor{mhold}{-1.91\%}  \\
                     & Test   Bullish  & 2023-10-01 & 2023-12-01 & 1671.00  & 2051.76  & \textcolor{minc}{22.59\%}  \\ \hline
\multirow{4}{*}{SOL} & Validation      & 2023-01-14 & 2023-03-12 & 18.29    & 18.24    & \textcolor{mhold}{-0.27\%}  \\
                     & Test   Bearish  & 2023-04-12 & 2023-06-16 & 23.02    & 14.76    & \textcolor{mdec}{-36.08\%} \\
                     & Test   Sideways & 2023-07-08 & 2023-08-31 & 21.49    & 20.83    & \textcolor{mhold}{-3.23\%}   \\
                     & Test   Bullish  & 2023-10-01 & 2023-12-01 & 21.39    & 59.25    & \textcolor{minc}{176.72\%} \\ \hline
\end{tabular}
\label{tab:period-stats}
\end{table*}


\stitle{Datasets.} To ensure our experiments are robust across different cryptocurrencies and market conditions, we base our study on a dataset covering several months, detailed in Table \ref{tab:period-stats}. This dataset reflects the recent market performance of BTC, ETH, and SOL, presenting challenges in capturing market trends and volatility. We divide the dataset into validation and test sets, using the former to select model hyperparameters and the latter to evaluate model performance. We carefully select the test period after September 2021, the GPT-3.5's knowledge cutoff date, to prevent data leakage. The dataset encompasses three market conditions: bull, sideways, and bear, allowing us to test the effectiveness of both the baselines and our model \cite{baroiu2023bitcoin, cagan2024stock}, ensuring reliable and robust experimental results.

\stitle{Evaluation Scheme.} We initialize the trading agent with 1 million US dollars, split equally between cash and BTC/ETH/SOL, to enable potential profits from both buying and selling cryptocurrencies. At the end of the trading session, we use the following widely-accepted metrics: \textbf{Return}, \textbf{Sharpe Ratio}, \textbf{Daily Return Mean}, and \textbf{Daily Return Std}. This evaluation scheme ensures a rigorous and unbiased assessment of both baseline strategies and our CryptoTrade agent.

\textbf{(1) Return} measures the overall performance of the trading strategy, calculated using the formula $\frac{w^{end}-w^{start}}{w^{start}}$, where $w^{start}$ and $w^{end}$ represent the starting and ending net worth, respectively.

\textbf{(2) Sharpe Ratio} assesses the risk-adjusted return, using the formula $\frac{\bar{r}-{r_f}}{\sigma}$, where $\bar{r}$ is the mean of daily returns, $\sigma$ is the standard deviation of daily returns, and $r_f$ is the risk-free return, set to 0 following SocioDojo \cite{cheng2024sociodojo}.

\textbf{(3) Daily Return Mean} is the average of the daily returns over the trading period, providing insight into the typical daily performance of the trading strategy.

\textbf{(4) Daily Return Std} is the standard deviation of the daily returns, indicating the volatility and risk associated with the daily performance of the trading strategy.

\stitle{Baseline Strategies.} To benchmark the performance of our CryptoTrade agent, we compare it against widely recognized baseline strategies in the trading domain. We present these baselines and hyperparameters in \autoref{baselines}.

\modify{
\subsection{Experimental Results}

The performance comparison presented in \autoref{tab:btc}, \autoref{tab:eth}, \autoref{tab:sol} between various trading strategies and our proposed CryptoTrade agent reveals significant insights into the efficacy of incorporating advanced data analysis techniques for cryptocurrency trading. The table highlights the returns and Sharpe Ratios for each method, where our CryptoTrade agent performs with outstanding percentage return and Sharpe Ratio. 
We outline the superiority of CryptoTrade in two key aspects:

\begin{table*}[h!]
    \centering
    \caption{Performance of each strategy on BTC under Bull, Sideways, and Bear market conditions. For each market condition and each metric, the best result is highlighted in bold text and the runner-up result is underlined.}
    \label{tab:btc}
    \vspace{0.2cm}
    \small 
    \resizebox{\textwidth}{!}{%
    \begin{tabular}{l|ccc|ccc|ccc}
        \toprule
        \textbf{Strategy} & \multicolumn{3}{c|}{\textbf{Total Return}} & \multicolumn{3}{c|}{\textbf{Daily Return}} & \multicolumn{3}{c}{\textbf{Sharpe Ratio}} \\
        \cmidrule(lr){2-4} \cmidrule(lr){5-7} \cmidrule(lr){8-10}
        & \textbf{Bull} & \textbf{Sideways} & \textbf{Bear} & \textbf{Bull} & \textbf{Sideways} & \textbf{Bear} & \textbf{Bull} & \textbf{Sideways} & \textbf{Bear} \\
        \midrule
        Buy and hold & \textbf{39.66} & \underline{-0.83} & -15.61 & \ms{\textbf{0.56}}{2.23} & \ms{\underline{0.00}}{1.74} & \ms{-0.24}{2.07} & \textbf{0.25} & \underline{0.00} & -0.11 \\
        SMA & 22.58 & \textbf{3.65} & -21.74 & \ms{0.35}{1.89} & \ms{\textbf{0.06}}{1.21} & \ms{-0.36}{1.25} & 0.18 & \textbf{0.05} & -0.29 \\
        SLMA & \underline{38.53} & -3.14 & \underline{-7.68} & \ms{\underline{0.55}}{2.21} & \ms{-0.04}{0.83} & \ms{\underline{-0.11}}{1.23} & \textbf{0.25} & -0.05 & -0.09 \\
        MACD & 13.57 & -6.71 & -9.51 & \ms{0.22}{1.45} & \ms{-0.09}{1.01} & \ms{-0.14}{1.56} & 0.15 & \underline{-0.09} & \underline{-0.09} \\
        Bollinger Bands & 2.97 & -3.19 & \textbf{-1.17} & \ms{0.05}{0.32} & \ms{-0.04}{0.87} & \ms{\textbf{-0.02}}{0.51} & 0.15 & -0.05 & \textbf{-0.03} \\
        \midrule
        LSTM & 31.67 & -4.13 & -17.20 & \ms{0.47}{2.11} & \ms{-0.05}{1.62} & \ms{-0.28}{1.27} & 0.22 & -0.03 & -0.22 \\
        Informer & 0.34 & -2.33 & -13.38 & \ms{0.01}{0.82} & \ms{-0.03}{0.54} & \ms{-0.21}{1.02} & 0.01 & -0.06 & -0.21 \\
        AutoFormer & 14.73 & -4.90 & -12.72 & \ms{0.24}{1.65} & \ms{-0.07}{1.15} & \ms{-0.20}{1.13} & 0.14 & -0.06 & -0.18 \\
        TimesNet & 2.84 & -5.12 & -13.64 & \ms{0.05}{1.06} & \ms{-0.07}{1.10} & \ms{-0.22}{1.04} & 0.05 & -0.06 & -0.21 \\
        PatchTST & 1.79 & -5.02 & -21.94 & \ms{0.03}{0.71} & \ms{-0.07}{0.57} & \ms{-0.37}{1.05} & 0.04 & -0.13 & -0.35 \\ \midrule
        Ours(GPT-4) & 26.35 & -4.07 & -11.72 & \ms{0.40}{1.76} & \ms{-0.05}{1.43} & \ms{-0.18}{1.67} & 0.23 & -0.04 & -0.11 \\    
        Ours(GPT-3.5-turbo) & 28.47 & -5.08 & -13.71 & \ms{0.43}{1.89} & \ms{-0.07}{1.14} & \ms{-0.21}{1.71} & \underline{0.23} & -0.06 & -0.12 \\         
        \bottomrule
    \end{tabular}
    }
\end{table*}

\begin{table*}[h!]
    \centering
    \caption{Performance of each strategy on ETH under Bull, Sideways, and Bear market conditions.}
    \vspace{0.2cm}
    \label{tab:eth}
    \small 
    \resizebox{\textwidth}{!}{%
    \begin{tabular}{l|ccc|ccc|ccc}
        \toprule
        \textbf{Strategy} & \multicolumn{3}{c|}{\textbf{Total Return (\%)}} & \multicolumn{3}{c|}{\textbf{Daily Return (\%)}} & \multicolumn{3}{c}{\textbf{Sharpe Ratio}} \\
        \cmidrule(lr){2-4} \cmidrule(lr){5-7} \cmidrule(lr){8-10}
        & \textbf{Bull} & \textbf{Sideways} & \textbf{Bear} & \textbf{Bull} & \textbf{Sideways} & \textbf{Bear} & \textbf{Bull} & \textbf{Sideways} & \textbf{Bear} \\
        \midrule
        Buy and Hold  & \underline{22.59} & -1.91 & -12.24 & \ms{\underline{0.36}}{2.62} & \ms{-0.01}{1.94} & \ms{-0.17}{2.39} & 0.14 & -0.00 & \underline{-0.07} \\
        SMA           & 10.17 & -5.45 & \underline{-10.12} & \ms{0.18}{2.29} & \ms{-0.15}{1.64} & \ms{\underline{-0.15}}{1.64} & 0.08 & -0.07 & -0.09 \\
        SLMA          & 5.20  & -2.62 & -15.90 & \ms{0.11}{2.37} & \ms{-0.03}{1.08} & \ms{-0.24}{1.86} & 0.05 & -0.03 & -0.13 \\
        MACD          & 7.72  & 0.77  & -12.15 & \ms{0.13}{1.22} & \ms{0.02}{1.43}  & \ms{-0.18}{1.56} & 0.10 & 0.01 & -0.12 \\
        Bollinger Bands & 2.59  & \textbf{4.47}  & \textbf{-0.41}  & \ms{0.04}{0.40} & \ms{\textbf{0.07}}{1.02}  & \ms{\textbf{0.00}}{0.58} & 0.11 & \underline{0.06} & \textbf{-0.01} \\
        \midrule
        LSTM          & 22.12 & \underline{1.27}  & -13.22 & \ms{0.36}{2.59} & \ms{\underline{0.02}}{1.11}  & \ms{-0.19}{2.36} & 0.14 & \textbf{0.15} & -0.08 \\
        Informer & 14.55 & -4.74 & -11.49 & \ms{0.23}{1.54} & \ms{-0.06}{1.45} & \ms{-0.17}{1.65} & \underline{0.15} & -0.04 & -0.10 \\
        AutoFormer & 7.77 & -10.06 & -19.44 & \ms{0.13}{1.81} & \ms{-0.14}{1.33} & \ms{-0.31}{1.61} & 0.08 & -0.10 & -0.20 \\
        TimesNet & 13.31 & -8.08 & -10.64 & \ms{0.21}{1.50} & \ms{-0.11}{1.08} & \ms{-0.16}{1.04} & 0.14 & -0.10 & -0.16 \\
        PatchTST & 8.95 & -9.64 & -13.76 & \ms{0.15}{1.37} & \ms{-0.13}{1.66} & \ms{-0.21}{1.39} & 0.11 & -0.11 & -0.15 \\ \midrule
        Ours(GPT-4) & \textbf{25.72} & 0.72 & -13.72 & \ms{\textbf{0.41}}{2.45} & \ms{0.03}{1.67} & \ms{-0.21}{2.02} & 0.17 & 0.02 & -0.10 \\ 
        Ours(GPT-3.5-turbo) & 18.91 & -5.02 & -14.40 & \ms{0.30}{2.01} & \ms{-0.06}{1.56} & \ms{-0.22}{2.08} & \textbf{0.15} & -0.04 & -0.10 \\ 
        \bottomrule
    \end{tabular}%
    }
\end{table*}

\begin{table*}[h!]
    \centering
    \caption{Performance of each strategy on SOL under Bull, Sideways, and Bear market conditions.}
    \vspace{0.2cm}
    \label{tab:sol}
    \small 
    \resizebox{\textwidth}{!}{%
    \begin{tabular}{l|ccc|ccc|ccc}
        \toprule
        \textbf{Strategy} & \multicolumn{3}{c|}{\textbf{Total Return (\%)}} & \multicolumn{3}{c|}{\textbf{Daily Return (\%)}} & \multicolumn{3}{c}{\textbf{Sharpe Ratio}} \\
        \cmidrule(lr){2-4} \cmidrule(lr){5-7} \cmidrule(lr){8-10}
        & \textbf{Bull} & \textbf{Sideways} & \textbf{Bear} & \textbf{Bull} & \textbf{Sideways} & \textbf{Bear} & \textbf{Bull} & \textbf{Sideways} & \textbf{Bear} \\
        \midrule
        Buy and Hold  & \textbf{176.72} & -3.23 & -36.08 & \ms{\textbf{1.83}}{6.00} & \ms{0.01}{3.92} & \ms{-0.61}{3.45} & \textbf{0.30} & 0.00 & -0.18 \\
        SMA           & 119.37 & -0.62 & \textbf{1.04}  & \ms{1.43}{5.67} & \ms{0.03}{3.06} & \ms{\textbf{0.02}}{0.10} & 0.25 & 0.01 & \textbf{0.16} \\
        SLMA          & \underline{169.98} & \textbf{6.22}  & \underline{-8.11} & \ms{\underline{1.78}}{5.93} & \ms{\textbf{0.16}}{3.23} & \ms{\underline{-0.11}}{1.88} & \textbf{0.30} & \textbf{0.05} & \underline{-0.06} \\
        MACD          & 23.25  & -9.78 & -21.07 & \ms{0.35}{1.76} & \ms{-0.16}{2.38} & \ms{-0.33}{2.44} & 0.20 & -0.07 & -0.13 \\
        Bollinger Bands & 2.92  & \underline{-0.46} & -21.69 & \ms{0.05}{0.35} & \ms{0.00}{1.23}  & \ms{-0.35}{1.75} & 0.13 & -0.00 & -0.20 \\
        \midrule
        LSTM          & 144.69 & -3.56 & -36.75 & \ms{1.61}{5.69} & \ms{\underline{0.01}}{3.90}  & \ms{-0.63}{3.43} & \underline{0.28} & 0.00 & -0.18 \\
        Informer & 41.85 & -6.55 & -26.13 & \ms{0.58}{1.90} & \ms{-0.10}{2.00} & \ms{-0.43}{2.36} & 0.31 & -0.05 & -0.18 \\
        AutoFormer & 35.86 & -6.17 & -23.56 & \ms{0.51}{1.97} & \ms{-0.10}{1.90} & \ms{-0.38}{2.35} & 0.26 & -0.05 & -0.16 \\
        TimesNet & 45.28 & -10.63 & -21.60 & \ms{0.64}{2.66} & \ms{-0.18}{2.01} & \ms{-0.35}{1.75} & 0.24 & -0.09 & -0.20 \\
        PatchTST & 18.45 & -7.10 & -27.86 & \ms{0.29}{1.57} & \ms{-0.11}{1.98} & \ms{-0.46}{2.49} & 0.18 & -0.06 & -0.19 \\ \midrule
        Ours(GPT-4) & 99.84 & -2.16 & -19.55 & \ms{1.24}{4.53} & \ms{0.01}{3.33} & \ms{-0.31}{2.35} & {0.27} & {0.00} & -0.13 \\ 
        Ours(GPT-3.5-turbo) & 102.45 & -13.05 & -24.08 & \ms{1.26}{4.54} & \ms{-0.23}{2.42} & \ms{-0.39}{2.60} & {0.28} & {-0.15} & -0.10 \\ 
        \bottomrule
    \end{tabular}
    }
\end{table*}


\stitle{Superior Performance under Different Market Conditions.} Remarkably, even without fine-tuning, CryptoTrade outperforms Transformer-based time-series baselines in most bases, demonstrating the robust capabilities of LLMs. Additionally, its performance is comparable to traditional trading signals like Buy and Hold and MACD, further validating the potential of LLM-based approaches. For instance, CryptoTrade (GPT-4o) excels in all metrics under ETH's bull market by 3\% in total return and sharpe ratio. While CryptoTrade (GPT-4o) may not always be the top performer in every scenario, it consistently surpasses more than half of the trading signals across different market conditions, even without fine-tuning. This highlights the effectiveness and versatility of CryptoTrade in leveraging LLMs to navigate the complexities of the cryptocurrency market.

\begin{figure*}
\centering 
\includegraphics[width=0.9\linewidth]{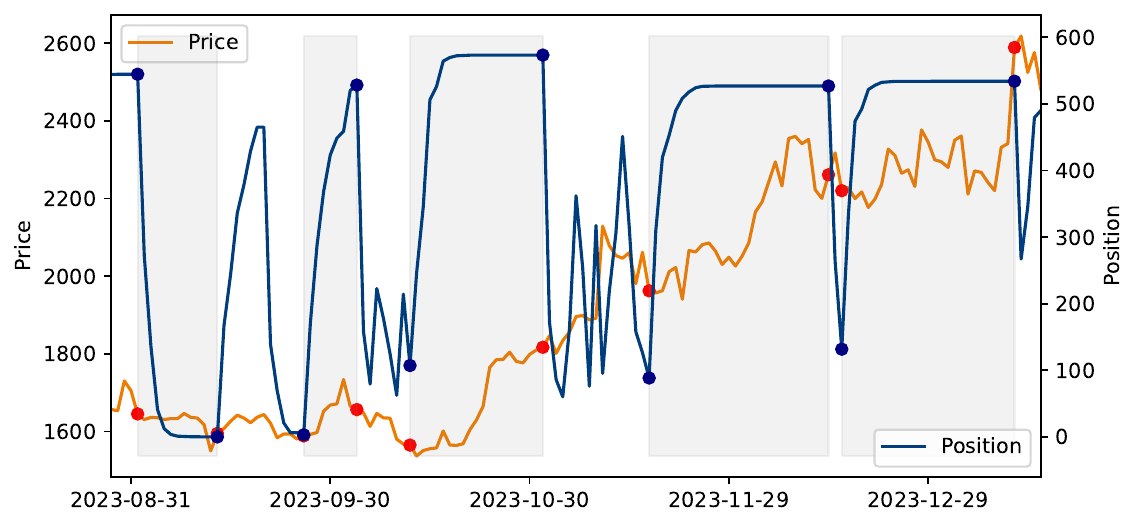} 
\caption{Significant profitable periods exploited by the CryptoTrade agent. The {yellow line} shows the daily opening prices of Ethereum in US dollars. The {blue line} tracks the daily positions, indicating the amount of Ethereum possessed on each day. The {blue dots} denote trading decisions when the agent largely alters its position by trading Ethereum. The {red dots} represent the corresponding trading prices. The agent successfully forecasts price changes, securing substantial profits through low-price purchases and high-price sales.} 
\label{fig:segments} 
\end{figure*}

\stitle{Successful Trend Predictions.} 
We draw \autoref{fig:segments} to demonstrate the correlation between Ethereum's opening prices and the positions held by the CryptoTrade agent, with the yellow and blue lines representing daily opening prices and Ethereum positions, respectively. The observed fluctuations highlight the market's volatility, while the alignment between position adjustments and price movements showcases the agent's proficiency in anticipating market trends. Unlike the static Buy and Hold strategy, CryptoTrade adopts a dynamic approach, optimizing trades based on market analysis—purchasing at lower prices and selling at peaks. This strategic adaptability, especially evident during shaded periods of preemptive position changes in anticipation of price shifts, underscores the agent's capacity for risk management and its adeptness at leveraging market volatility for profit, marking a significant advancement over traditional trading strategies.

\subsection{Ablation Study}
The ablation study presented in \autoref{tab:ablation} critically examines the individual components of the prompt used by the CryptoTrade (GPT-4o) agent. By systematically removing key elements from the full prompt and observing the impact on percentage return and Sharpe ratio during a bull market for ETH, we can identify the contribution of each component to the overall performance of the trading strategy.We highlight the following two insights from the results:

\begin{table}[!ht]
\centering
\caption{Ablation study on prompt components of the CryptoTrade agent. Base prompt encompasses necessary context including trading rules, valid action space, current cash and ETH holdings, and recent ETH prices.}
\vspace{0.2cm}
\begin{tabular}{ccc}
\hline
Prompt Components & Return (\%) & Sharpe Ratio \\ \hline
Full              & \textbf{28.47}      & \textbf{0.23}        \\
w/o Reflection    & 17.14       & 0.06        \\
w/o News          & 19.69       & 0.06         \\
w/o TxnStats      & 12.70       & 0.05         \\
w/o Technical     & 17.27       & 0.05         \\
Base              & 8.40        & 0.03         \\ \hline
\end{tabular}
\label{tab:ablation}
\end{table}


\stitle{Superiority of the Full Prompt.} 
The full prompt significantly outshines all other configurations with reduced components. The advantage of employing a full prompt over all deducted variants is rooted in the integration of diverse data sources. The full prompt encompasses the comprehensive price data, news analysis, technical indicators, on-chain transaction statistics, and reflective analysis to offer a holistic view of the market. This comprehensive approach allows the CryptoTrade agent to leverage a wide array of information, enabling it to navigate the complexities of the cryptocurrency market with more nuanced and informed trading decisions.

\stitle{Advantage of Crypto Transaction Statistics.} The omission of Ethereum transaction statistics results in a significant decrease of the outcome by around 16\%, underscoring the indispensable role of on-chain statistics in enhancing trading strategies. This observation highlights the necessity of integrating on-chain transaction data, revealing its unique value in enriching the decision-making process in the cryptocurrency trading tasks.



\subsection{Case Study}

\begin{figure*}[!ht]
\centering 
\vspace{0.2cm}
\includegraphics[width=0.9\linewidth]{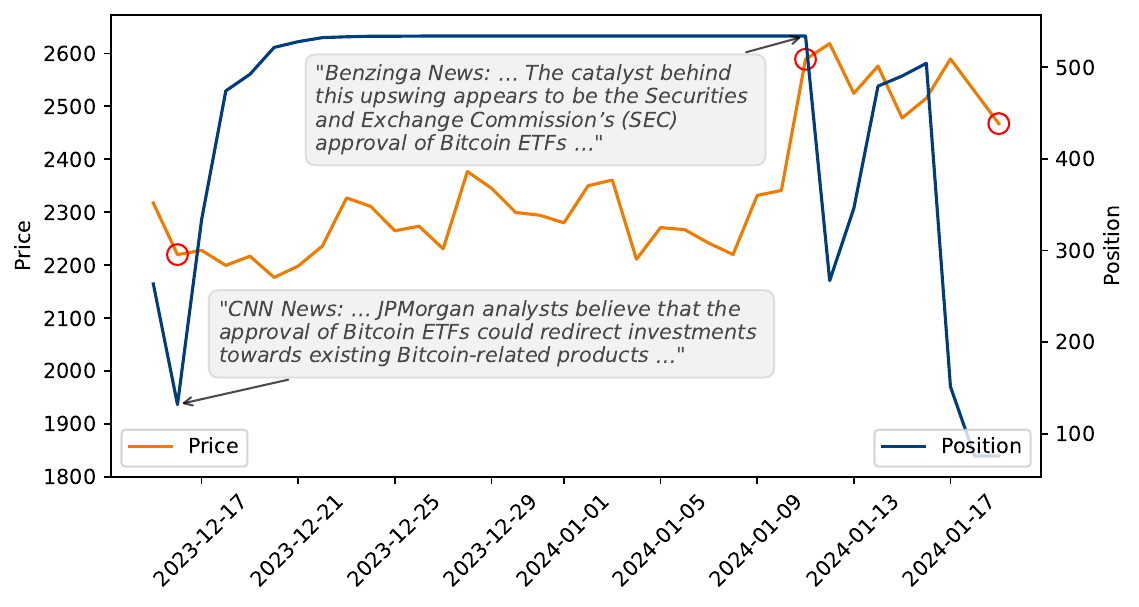} 
\caption{Case study of CryptoTrade's actions in response to news reports on early rumor and the actual event of Bitcoin ETF approval, which takes place on Jan 11, 2024. The red circles denote the trading prices. The agent successfully benefits from a "buy the rumor, sell the news" strategy. } 
\label{fig:case} 
\end{figure*}
}

To assess the adaptability and responsiveness of the CryptoTrade agent, we conduct a case study focusing on its responsive actions in the context of the cryptocurrency market's major events, illustrated in \autoref{fig:case}. It reveals that CryptoTrade's strategy aligns with the "buy the rumor, sell the news" principle, effectively capitalizing on early signs of the Bitcoin ETF approval event, a scenario known to trigger market rallies due to speculative trading. By entering the market early, CryptoTrade secures positions at lower costs ahead of the rally.


As the approval of the Bitcoin ETF becomes a reality, the sentiment reaches a crescendo, resulting in inflated asset prices due to heightened demand. CryptoTrade, adhering to its strategic motivation, takes this peak as an optimal point to sell, which is validated in the subsequent decline in the Ethereum price. This strategic exit allows CryptoTrade to realize gains before the market adjusted to the new equilibrium, which results in a price pullback as early speculators take profits and the market sentiment normalizes.

To sum up, CryptoTrade's provident actions underscore the delicate balance between foresight and timing in trading strategies. This case study demonstrates that an informed and timely response to market signals — both rumors and confirmed news — can yield advantageous outcomes. It also highlights the CryptoTrade agent's understanding of market psychology and its ability to translate this into profitable trading decisions.


\section{Related Work}
\textbf{LLMs for Economics and Financial Decisions} 
Recent advancements in LLMs have significantly influenced economics and financial decision-making. Specialized LLMs like FinGPT, BloombergGPT, FinMA \cite{liu2023fingpt, wu2023bloomberggpt, xie2023pixiu} are tailored for finance, handling tasks such as sentiment analysis, entity recognition, and question-answering. Another research direction uses LLMs for financial time-series forecasting. A notable contribution by \cite{yu2023temporal} employed zero-shot or few-shot inference with GPT-4 and instruction-based fine-tuning with LlaMA to enhance cross-sequence reasoning and multi-modal signal integration. Additionally, the development of LLM-based agents for financial trading has gained attention. Sociodojo \cite{cheng2024sociodojo} created analytical agents for stock portfolio management, showing the potential for generating "hyperportfolios." Despite these advancements, the focus has largely been on the stock market \cite{koa2024learning, chen2023chatgpt}, leaving a gap in the exploration of the cryptocurrency market where the on-chain data is approachable and with much information. Our work aims to address this gap by leveraging both on-chain and off-chain data to navigate the dynamic cryptocurrency market.

\textbf{Time-Series Forecasting for Financial Markets} Time-series forecasting has long been a cornerstone of research in economics and financial markets. Early studies focused on predicting stock market prices using methodologies such as machine learning \cite{leung2021promises, patel2014stock}, reinforcement learning \cite{lee2001stock}, and traditional time-series models \cite{herwartz2017stock}. The Long Short-Term Memory (LSTM) model has emerged as particularly influential \cite{sunny2020deep} for its capability to process and analyze time-series data. With the rise of blockchain technology and cryptocurrencies, these techniques have been extended to crypto assets \cite{khedr2021cryptocurrency}. Recent research has evaluated the impact of various predictors on cryptocurrency pricing and returns, using both on-chain data—such as historical transactions and market volume \cite{ferdiansyah2019lstm}—and off-chain factors like social media trends and news sentiment \cite{abraham2018cryptocurrency, pang2019cryptocurrency}. These studies underscore the effectiveness of integrating diverse data sources for forecasting the volatile dynamics of the cryptocurrency market. Apart from above, Transformer-based models have shown particular promise in this area, with state-of-the-art models like Informer \cite{zhou2021informer}, AutoFormer \cite{wu2021autoformer}, PatchTST \cite{nie2022time}, and TimesNet \cite{wu2022timesnet} further advancing time-series forecasting.


\textbf{Self-Reflective Language Agents}
The Self-Refine framework introduces an advanced approach for autonomous advancement through self-evaluation and iterative self-improvement \cite{madaan2024self}. This approach, along with efforts to automatically refine prompts \cite{pryzant2023automatic, ye2024investigating} and provide automated feedback to enhance reasoning capabilities \cite{paul2023refiner}, marks significant progress in the field. Notably, the "Reflexion" framework by \cite{shinn2024reflexion} revolutionizes the reinforcement of language agents by utilizing linguistic feedback and reflective text within an episodic memory buffer, diverging from traditional weight update methods. These advancements highlight the potential for LLMs to learn from their errors and evolve through self-reflection. Despite these developments, there is still untapped potential in applying self-reflective language agents to financial decision-making, particularly in cryptocurrency markets. This work aims to bridge that gap by investigating the application of self-reflective mechanisms to enhance financial decision-making processes in cryptocurrency trading.
\section{Conclusion}
We propose the CryptoTrade agent, an innovative approach to cryptocurrency trading that leverages advanced data analysis and LLMs. By integrating both on-chain and off-chain data, along with a self-reflective component, the CryptoTrade agent demonstrates a sophisticated understanding of market dynamics and achieves relatively high returns in cryptocurrency trading. Our comprehensive experiments comparing the CryptoTrade agent to traditional trading strategies and time-series models reveal its superior ability to navigate the volatile cryptocurrency market, consistently achieving relatively high returns on investment under different market conditions. This research underscores the significant potential of LLM-driven strategies in enhancing trading performance and sets a new benchmark for cryptocurrency trading with LLMs. 

\section{Limitations}
One limitation of the current CryptoTrade framework is the reliance on a relatively limited dataset. To address this, we plan to enrich the dataset with additional off-chain data. Another limitation is the frequency of trading actions, which is currently set to day-to-day. We aim to refine this to hour-to-hour or minute-to-minute intervals to further optimize returns in the cryptocurrency market. Additionally, we have identified that the lack of fine-tuning for the LLMs using the validation set may be a significant factor behind the LLM-based agents' underperformance compared to traditional trading signals. To improve the reliability of our forecasts, we intend to fine-tune the LLMs with the validation set.

\section{Broader Impact}
One potential broader impact of our research is the risk that individuals may follow the trading strategies we provide and subsequently incur financial losses. It is important to emphasize that these strategies are intended for academic research only. CryptoTrade is not for investment recommendations.

\section*{Acknowledgements}
This work originated as a course project in CS6216 (Advanced Topics in Machine Learning, Spring 2024) taught by Prof. Qizhe Xie at the National University of Singapore.


\bibliographystyle{acl_natbib}
\bibliography{bibliography}

@article{wu2023bloomberggpt,
	title        = {Bloomberggpt: A large language model for finance},
	author       = {Wu, Shijie and Irsoy, Ozan and Lu, Steven and Dabravolski, Vadim and Dredze, Mark and Gehrmann, Sebastian and Kambadur, Prabhanjan and Rosenberg, David and Mann, Gideon},
	year         = 2023,
	journal      = {arXiv preprint arXiv:2303.17564}
}

@book{cagan2024stock,
  title={Stock Market 101: From Bull and Bear Markets to Dividends, Shares, and Margins—Your Essential Guide to the Stock Market},
  author={Cagan, Michele},
  year={2024},
  publisher={Simon and Schuster}
}

@article{baroiu2023bitcoin,
  title={Bitcoin volatility in bull vs. bear market-insights from analyzing on-chain metrics and Twitter posts},
  author={Baroiu, Alexandru Costin and Diaconita, Vlad and Oprea, Simona Vasilica},
  journal={PeerJ Computer Science},
  volume={9},
  pages={e1750},
  year={2023},
  publisher={PeerJ Inc.}
}

@inproceedings{ye2024investigating,
  title={Investigating the effectiveness of task-agnostic prefix prompt for instruction following},
  author={Ye, Seonghyeon and Hwang, Hyeonbin and Yang, Sohee and Yun, Hyeongu and Kim, Yireun and Seo, Minjoon},
  booktitle={Proceedings of the AAAI Conference on Artificial Intelligence},
  volume={38},
  number={17},
  pages={19386--19394},
  year={2024}
}

@article{ren2023interoperability,
  title={Interoperability in blockchain: A survey},
  author={Ren, Kunpeng and Ho, Nhut-Minh and Loghin, Dumitrel and Nguyen, Thanh-Toan and Ooi, Beng Chin and Ta, Quang-Trung and Zhu, Feida},
  journal={IEEE Transactions on Knowledge and Data Engineering},
  year={2023},
  publisher={IEEE}
}

@inproceedings{feichtinger2023hidden,
  title={The hidden shortcomings of (d) aos--an empirical study of on-chain governance},
  author={Feichtinger, Rainer and Fritsch, Robin and Vonlanthen, Yann and Wattenhofer, Roger},
  booktitle={International Conference on Financial Cryptography and Data Security},
  pages={165--185},
  year={2023},
  organization={Springer}
}

@article{wei2023cryptocurrency,
  title={Cryptocurrency uncertainty and volatility forecasting of precious metal futures markets},
  author={Wei, Yu and Wang, Yizhi and Lucey, Brian M and Vigne, Samuel A},
  journal={Journal of Commodity Markets},
  volume={29},
  pages={100305},
  year={2023},
  publisher={Elsevier}
}

@article{drozdz2023mature,
  title={What is mature and what is still emerging in the cryptocurrency market?},
  author={Dro{\.z}d{\.z}, Stanis{\l}aw and Kwapie{\'n}, Jaros{\l}aw and W{\k{a}}torek, Marcin},
  journal={Entropy},
  volume={25},
  number={5},
  pages={772},
  year={2023},
  publisher={MDPI}
}

@article{pu2023summarization,
  title={Summarization is (almost) dead},
  author={Pu, Xiao and Gao, Mingqi and Wan, Xiaojun},
  journal={arXiv preprint arXiv:2309.09558},
  year={2023}
}

@article{liang2022holistic,
  title={Holistic evaluation of language models},
  author={Liang, Percy and Bommasani, Rishi and Lee, Tony and Tsipras, Dimitris and Soylu, Dilara and Yasunaga, Michihiro and Zhang, Yian and Narayanan, Deepak and Wu, Yuhuai and Kumar, Ananya and others},
  journal={arXiv preprint arXiv:2211.09110},
  year={2022}
}

@article{yang2024harnessing,
  title={Harnessing the power of llms in practice: A survey on chatgpt and beyond},
  author={Yang, Jingfeng and Jin, Hongye and Tang, Ruixiang and Han, Xiaotian and Feng, Qizhang and Jiang, Haoming and Zhong, Shaochen and Yin, Bing and Hu, Xia},
  journal={ACM Transactions on Knowledge Discovery from Data},
  volume={18},
  number={6},
  pages={1--32},
  year={2024},
  publisher={ACM New York, NY}
}

@article{chen2023chatgpt,
  title={ChatGPT informed graph neural network for stock movement prediction},
  author={Chen, Zihan and Zheng, Lei Nico and Lu, Cheng and Yuan, Jialu and Zhu, Di},
  journal={arXiv preprint arXiv:2306.03763},
  year={2023}
}

@article{xie2023pixiu,
  title={Pixiu: A large language model, instruction data and evaluation benchmark for finance},
  author={Xie, Qianqian and Han, Weiguang and Zhang, Xiao and Lai, Yanzhao and Peng, Min and Lopez-Lira, Alejandro and Huang, Jimin},
  journal={arXiv preprint arXiv:2306.05443},
  year={2023}
}

@inproceedings{koa2024learning,
  title={Learning to Generate Explainable Stock Predictions using Self-Reflective Large Language Models},
  author={Koa, Kelvin JL and Ma, Yunshan and Ng, Ritchie and Chua, Tat-Seng},
  booktitle={Proceedings of the ACM on Web Conference 2024},
  pages={4304--4315},
  year={2024}
}

@article{zhao2023survey,
  title={A survey of large language models},
  author={Zhao, Wayne Xin and Zhou, Kun and Li, Junyi and Tang, Tianyi and Wang, Xiaolei and Hou, Yupeng and Min, Yingqian and Zhang, Beichen and Zhang, Junjie and Dong, Zican and others},
  journal={arXiv preprint arXiv:2303.18223},
  year={2023}
}

@article{zhang2023multimodal,
  title={Multimodal chain-of-thought reasoning in language models},
  author={Zhang, Zhuosheng and Zhang, Aston and Li, Mu and Zhao, Hai and Karypis, George and Smola, Alex},
  journal={arXiv preprint arXiv:2302.00923},
  year={2023}
}

@article{wei2022chain,
  title={Chain-of-thought prompting elicits reasoning in large language models},
  author={Wei, Jason and Wang, Xuezhi and Schuurmans, Dale and Bosma, Maarten and Xia, Fei and Chi, Ed and Le, Quoc V and Zhou, Denny and others},
  journal={Advances in neural information processing systems},
  volume={35},
  pages={24824--24837},
  year={2022}
}

@inproceedings{wu2022timesnet,
  title={Timesnet: Temporal 2d-variation modeling for general time series analysis},
  author={Wu, Haixu and Hu, Tengge and Liu, Yong and Zhou, Hang and Wang, Jianmin and Long, Mingsheng},
  booktitle={The eleventh international conference on learning representations},
  year={2022}
}

@article{yi2024frequency,
  title={Frequency-domain MLPs are more effective learners in time series forecasting},
  author={Yi, Kun and Zhang, Qi and Fan, Wei and Wang, Shoujin and Wang, Pengyang and He, Hui and An, Ning and Lian, Defu and Cao, Longbing and Niu, Zhendong},
  journal={Advances in Neural Information Processing Systems},
  volume={36},
  year={2024}
}

@article{nie2022time,
  title={A time series is worth 64 words: Long-term forecasting with transformers},
  author={Nie, Yuqi and Nguyen, Nam H and Sinthong, Phanwadee and Kalagnanam, Jayant},
  journal={arXiv preprint arXiv:2211.14730},
  year={2022}
}

@article{wu2021autoformer,
  title={Autoformer: Decomposition transformers with auto-correlation for long-term series forecasting},
  author={Wu, Haixu and Xu, Jiehui and Wang, Jianmin and Long, Mingsheng},
  journal={Advances in neural information processing systems},
  volume={34},
  pages={22419--22430},
  year={2021}
}

@inproceedings{zhou2021informer,
  title={Informer: Beyond efficient transformer for long sequence time-series forecasting},
  author={Zhou, Haoyi and Zhang, Shanghang and Peng, Jieqi and Zhang, Shuai and Li, Jianxin and Xiong, Hui and Zhang, Wancai},
  booktitle={Proceedings of the AAAI conference on artificial intelligence},
  volume={35},
  number={12},
  pages={11106--11115},
  year={2021}
}

@article{gencay1996non,
  title={Non-linear prediction of security returns with moving average rules},
  author={Gencay, Ramazan},
  journal={Journal of Forecasting},
  volume={15},
  number={3},
  pages={165--174},
  year={1996},
  publisher={Wiley Online Library}
}

@article{wang2018predicting,
  title={Predicting stock price trend using MACD optimized by historical volatility},
  author={Wang, Jian and Kim, Junseok},
  journal={Mathematical Problems in Engineering},
  volume={2018},
  pages={1--12},
  year={2018},
  publisher={Hindawi Limited}
}

@article{day2023profitability,
  title={The profitability of Bollinger Bands trading bitcoin futures},
  author={Day, Min-Yuh and Cheng, Yirung and Huang, Paoyu and Ni, Yensen},
  journal={Applied Economics Letters},
  volume={30},
  number={11},
  pages={1437--1443},
  year={2023},
  publisher={Taylor \& Francis}
}

@article{yu2023temporal,
	title        = {Temporal Data Meets LLM--Explainable Financial Time Series Forecasting},
	author       = {Yu, Xinli and Chen, Zheng and Ling, Yuan and Dong, Shujing and Liu, Zongyi and Lu, Yanbin},
	year         = 2023,
	journal      = {arXiv preprint arXiv:2306.11025}
}

@article{liu2023fingpt,
	title        = {Fingpt: Democratizing internet-scale data for financial large language models},
	author       = {Liu, Xiao-Yang and Wang, Guoxuan and Zha, Daochen},
	year         = 2023,
	journal      = {arXiv preprint arXiv:2307.10485}
}

@inproceedings{cheng2024sociodojo,
	title        = {SocioDojo: Building Lifelong Analytical Agents with Real-world Text and Time Series},
	author       = {Junyan Cheng and Peter Chin},
	year         = 2024,
	booktitle    = {The Twelfth International Conference on Learning Representations},
	url          = {https://openreview.net/forum?id=s9z0HzWJJp}
}

@article{khedr2021cryptocurrency,
	title        = {Cryptocurrency price prediction using traditional statistical and machine-learning techniques: A survey},
	author       = {Khedr, Ahmed M and Arif, Ifra and El-Bannany, Magdi and Alhashmi, Saadat M and Sreedharan, Meenu},
	year         = 2021,
	journal      = {Intelligent Systems in Accounting, Finance and Management},
	publisher    = {Wiley Online Library},
	volume       = 28,
	number       = 1,
	pages        = {3--34}
}

@article{leung2021promises,
	title        = {The promises and pitfalls of machine learning for predicting stock returns},
	author       = {Leung, Edward and Lohre, Harald and Mischlich, David and Shea, Yifei and Stroh, Maximilian},
	year         = 2021,
	journal      = {The Journal of Financial Data Science},
	publisher    = {Institutional Investor Journals Umbrella}
}

@inproceedings{lee2001stock,
	title        = {Stock price prediction using reinforcement learning},
	author       = {Lee, Jae Won},
	year         = 2001,
	booktitle    = {ISIE 2001. 2001 IEEE International Symposium on Industrial Electronics Proceedings (Cat. No. 01TH8570)},
	volume       = 1,
	pages        = {690--695},
	organization = {IEEE}
}

@article{herwartz2017stock,
	title        = {Stock return prediction under GARCH—An empirical assessment},
	author       = {Herwartz, Helmut},
	year         = 2017,
	journal      = {International Journal of Forecasting},
	publisher    = {Elsevier},
	volume       = 33,
	number       = 3,
	pages        = {569--580}
}

@article{patel2014stock,
	title        = {Stock price prediction using artificial neural network},
	author       = {Patel, Mayankkumar B and Yalamalle, Sunil R},
	year         = 2014,
	journal      = {International Journal of Innovative Research in Science, Engineering and Technology},
	publisher    = {Citeseer},
	volume       = 3,
	number       = 6,
	pages        = {13755--13762}
}

@inproceedings{sunny2020deep,
	title        = {Deep learning-based stock price prediction using LSTM and bi-directional LSTM model},
	author       = {Sunny, Md Arif Istiake and Maswood, Mirza Mohd Shahriar and Alharbi, Abdullah G},
	year         = 2020,
	booktitle    = {2020 2nd novel intelligent and leading emerging sciences conference (NILES)},
	pages        = {87--92},
	organization = {IEEE}
}

@article{abraham2018cryptocurrency,
	title        = {Cryptocurrency price prediction using tweet volumes and sentiment analysis},
	author       = {Abraham, Jethin and Higdon, Daniel and Nelson, John and Ibarra, Juan},
	year         = 2018,
	journal      = {SMU Data Science Review},
	volume       = 1,
	number       = 3,
	pages        = 1
}

@inproceedings{pang2019cryptocurrency,
	title        = {Cryptocurrency price prediction using time series and social sentiment data},
	author       = {Pang, Yan and Sundararaj, Ganeshkumar and Ren, Jiewen},
	year         = 2019,
	booktitle    = {Proceedings of the 6th IEEE/ACM International Conference on Big Data Computing, Applications and Technologies},
	pages        = {35--41}
}

@inproceedings{ferdiansyah2019lstm,
	title        = {A lstm-method for bitcoin price prediction: A case study yahoo finance stock market},
	author       = {Ferdiansyah, Ferdiansyah and Othman, Siti Hajar and Radzi, Raja Zahilah Raja Md and Stiawan, Deris and Sazaki, Yoppy and Ependi, Usman},
	year         = 2019,
	booktitle    = {2019 international conference on electrical engineering and computer science (ICECOS)},
	pages        = {206--210},
	organization = {IEEE}
}

@article{shinn2024reflexion,
	title        = {Reflexion: Language agents with verbal reinforcement learning},
	author       = {Shinn, Noah and Cassano, Federico and Gopinath, Ashwin and Narasimhan, Karthik and Yao, Shunyu},
	year         = 2024,
	journal      = {Advances in Neural Information Processing Systems},
	volume       = 36
}

@article{madaan2024self,
	title        = {Self-refine: Iterative refinement with self-feedback},
	author       = {Madaan, Aman and Tandon, Niket and Gupta, Prakhar and Hallinan, Skyler and Gao, Luyu and Wiegreffe, Sarah and Alon, Uri and Dziri, Nouha and Prabhumoye, Shrimai and Yang, Yiming and others},
	year         = 2024,
	journal      = {Advances in Neural Information Processing Systems},
	volume       = 36
}

@article{pryzant2023automatic,
	title        = {Automatic prompt optimization with" gradient descent" and beam search},
	author       = {Pryzant, Reid and Iter, Dan and Li, Jerry and Lee, Yin Tat and Zhu, Chenguang and Zeng, Michael},
	year         = 2023,
	journal      = {arXiv preprint arXiv:2305.03495}
}

@article{paul2023refiner,
	title        = {Refiner: Reasoning feedback on intermediate representations},
	author       = {Paul, Debjit and Ismayilzada, Mete and Peyrard, Maxime and Borges, Beatriz and Bosselut, Antoine and West, Robert and Faltings, Boi},
	year         = 2023,
	journal      = {arXiv preprint arXiv:2304.01904}
}

@book{narayanan2016bitcoin,
  title={Bitcoin and cryptocurrency technologies: a comprehensive introduction},
  author={Narayanan, Arvind and Bonneau, Joseph and Felten, Edward and Miller, Andrew and Goldfeder, Steven},
  year={2016},
  publisher={Princeton University Press}
}


\clearpage
\appendix
\section*{Appendix}
\section{License} \label{license}

The CryptoTrade's dataset is released under the Creative Commons Attribution-NonCommercial-ShareAlike (CC BY-NC-SA) license. This means that anyone can use, distribute, and modify the data for non-commercial purposes as long as they give proper attribution and share the derivative works under the same license terms.

\section{Data Ethics} \label{collection.ethics}

\subsection{On-chain Data}
We collect on-chain data from CoinMarketCap\footnote{\url{https://coinmarketcap.com}} and Dune\footnote{\url{https://dune.com/home}}. According to CoinMarketCap's Terms of Service\footnote{\url{https://coinmarketcap.com/terms/}}, we are granted a limited, personal, non-exclusive, non-sub-licensable, and non-transferable license to use the content and service solely for personal use. We agree not to use the service or any of the content for any commercial purpose, and we adhere to these requirements. Regarding Dune's Terms of Service\footnote{\url{https://dune.com/terms}}, we are permitted to access Dune’s application programming interfaces (the “API”) to perform SQL queries on blockchain data.

\subsection{Off-chain News}
We employ the Gnews\footnote{\url{https://pypi.org/project/gnews/}} to systematically gather news articles related to each cryptocurrency. According to Gnews' Terms of Service\footnote{\url{https://gnews.io/terms/}}, we can download the news for non-commercial transitory viewing only, and we cannot modify or copy the materials, use the materials for any commercial purpose or any public display, attempt to reverse engineer any software contained on Gnews API's website, remove any copyright or other proprietary notations from the materials, or transfer the materials to another person or "mirror" the materials on any other server. We adhere to these conditions in our CryptoTrade dataset. 


\section{Experimental Environment} \label{env}
All models in our experiments were implemented using Pytorch 2.0.0 in Python 3.9.16, and run on a robust Linux workstation. This system is equipped with two Intel(R) Xeon(R) Gold 6226R CPUs, each operating at a base frequency of 2.90 GHz and a max turbo frequency of 3.90 GHz. With 16 cores each, capable of supporting 32 threads, these CPUs offer a total of 64 logical CPUs for efficient multitasking and parallel computing. The workstation is further complemented by a potent GPU setup, comprising eight NVIDIA GeForce RTX 3090 GPUs, each providing 24.576 GB of memory. The operation of these GPUs is managed by the NVIDIA-SMI 525.60.13 driver and CUDA 12.0, ensuring optimal computational performance for our tasks.

\section{Analysts Examples} \label{analysts}
In this section, we provide some examples of News Analyst, Market Analyst, Reflection Analyst, and Trading Analyst.

\begin{figure}[!ht]
\centering 
\includegraphics[width=1\linewidth]{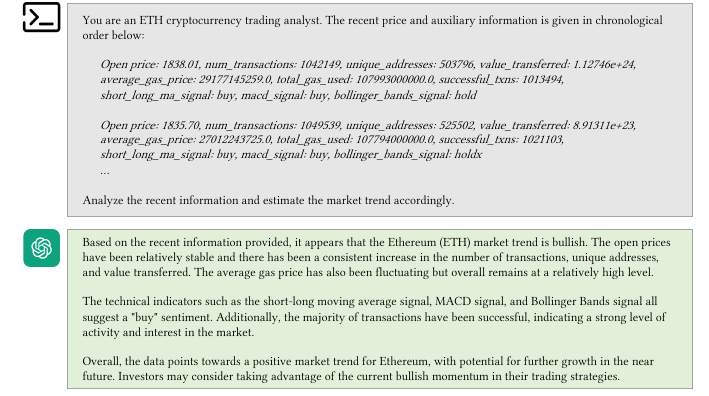} 
\caption{A sample of the Market Analyst.} 
\label{fig:market} 
\end{figure}

\begin{figure}[!ht]
\centering 
\includegraphics[width=1\linewidth]{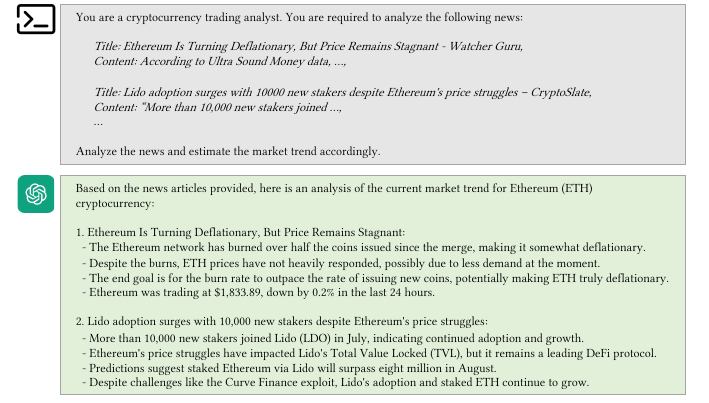} 
\caption{A sample of the News Analyst.} 
\label{fig:news} 
\end{figure}

\begin{figure}[!ht]
\centering 
\includegraphics[width=1\linewidth]{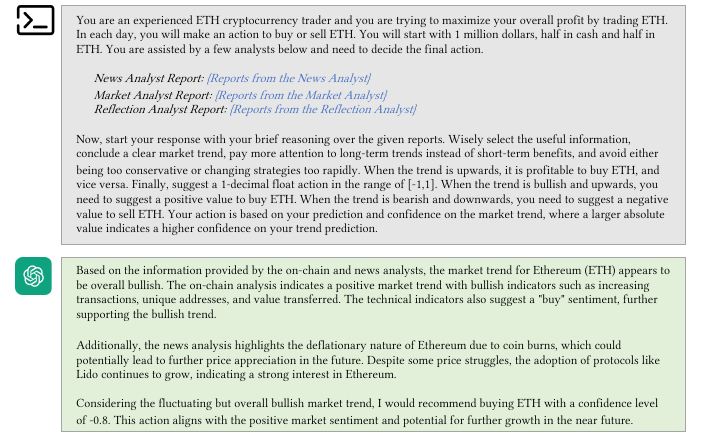} 
\caption{A sample of the Trading Analyst. }
\label{fig:investment} 
\end{figure}

\begin{figure}[!ht]
\centering 
\includegraphics[width=1\linewidth]{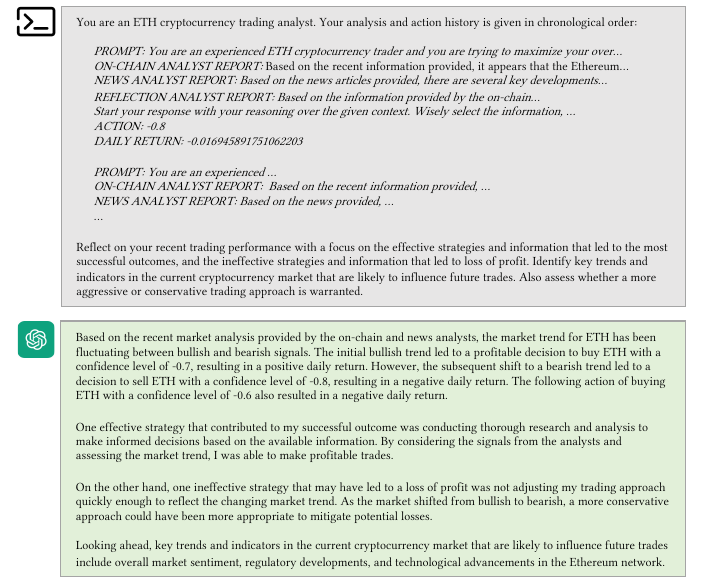} 
\caption{A sample of the Reflection Analyst.} 
\label{fig:reflection} 
\end{figure}


\section{Baselines} \label{baselines}
\begin{enumerate}
    \item \textbf{Buy and Hold:} A straightforward strategy where an asset is purchased at the beginning of the period and held until its end.
    
    \item \textbf{SMA \cite{gencay1996non}:} SMA triggers buy or sell decisions based on the asset's price relative to its moving average. We finetune the SMA period by testing different window sizes $[5, 10, 15, 20, 30]$. The optimal period is selected based on the best performance on the validation set.
    
    \item \textbf{SLMA \cite{wang2018predicting}:} SLMA involves two moving averages of different lengths, with trading signals generated at their crossover points. We use different combinations of short and long SMA periods, selecting the optimal ones based on validation set performance.
    
    \item \textbf{MACD \cite{wang2018predicting}:} A strategy that uses the MACD indicator to identify potential buy and sell opportunities based on the momentum of the asset. The MACD is calculated as the difference between the 12-day EMA and the 26-day EMA, with a 9-day EMA of the MACD line serving as the signal line. EMA stands for Exponential Moving Average. It is a type of moving average that places a greater weight and significance on the most recent data points.
    
    \item \textbf{Bollinger Bands \cite{day2023profitability}:} This strategy generates trading signals based on price movements relative to the middle, lower, and upper Bollinger Bands. Bollinger Bands are constructed using a 20-day SMA and a multiplier (commonly set to 2) for the standard deviation. We use the recommended period and multiplier settings for this strategy.

    \item \textbf{LSTM \cite{ferdiansyah2019lstm}):} This strategy involves comparing today's price with the predicted price for tomorrow to identify potential buying and selling opportunities. We fine-tune the look-back window size using values in $[1, 3, 5, 10, 20, 30]$ and select the parameters that perform best on the validation set.
    
    \item \textbf{Informer \cite{zhou2021informer}:} Informer utilizes an efficient self-attention mechanism to capture dependencies among variables. We adopt the recommended configuration for our experimental settings: a dropout rate of 0.05, two encoder layers, one decoder layer, a learning rate of 0.0001, and the Adam optimizer \cite{yi2024frequency}. The look-back window size is selected using the same procedure as for the LSTM.

    \item \textbf{AutoFormer \cite{wu2021autoformer}:} AutoFormer introduces a decomposition architecture by embedding the series decomposition block as an inner operator, allowing for the progressive aggregation of the long-term trend from intermediate predictions. We use the recommended configuration for our experimental settings \cite{yi2024frequency}. The look-back window size is selected using the same procedure as for the LSTM.
    
    \item \textbf{TimesNet \cite{wu2022timesnet}:} TimesNet provides a general framework for various time-series forecasting tasks. We adopt the recommended configurations for our experimental settings \cite{wu2022timesnet}. The look-back window size is selected using the same procedure as for the LSTM.
    
    \item \textbf{PatchTST \cite{nie2022time}:} PatchTST proposes an effective design for Transformer-based models in time series forecasting by introducing two key components: patching and a channel-independent structure \cite{yi2024frequency}. The recommended configurations are used for our experimental settings. The look-back window size is selected using the same procedure as for the LSTM.
\end{enumerate}

\section{Author Statement}
As authors of the CryptoTrade, we hereby declare that we assume full responsibility for any liability or infringement of third-party rights that may come up from the use of our data. We confirm that we have obtained all necessary permissions and/or licenses needed to share this data with others for their own use. In doing so, we agree to indemnify and hold harmless any person or entity that may suffer damages resulting from our actions.

Furthermore, we confirm that our CryptoTrade dataset is released under the Creative Commons Attribution-NonCommercial-ShareAlike (CC BY-NC-SA) license. This license allows anyone to use, distribute, and modify our data for non-commercial purposes as long as they give proper attribution and share the derivative works under the same license terms. We believe that this licensing model aligns with our goal of promoting open access to high-quality data while respecting the intellectual property rights of all parties involved.

\section{Hosting Plan}
After careful consideration, we have chosen to host our code and data on GitHub. Our decision is based on various factors, including the platform's ease of use, cost-effectiveness, and scalability. We understand that accessibility is key when it comes to data management, which is why we will ensure that our data is easily accessible through a curated interface. We also recognize the importance of maintaining the platform's stability and functionality, and as such, we will provide the necessary maintenance to ensure that it remains up-to-date, bug-free, and running smoothly.

At the heart of our project is the belief in open access to data, and we are committed to making our data available to those who need it. As part of this commitment, we will be updating our GitHub repository regularly, so that users can rely on timely access to the most current information. We hope that by using GitHub as our hosting platform, we can provide a user-friendly and reliable solution for sharing our data with others.

\clearpage

\end{document}